\documentclass[aps,prl,twocolumn,twoside,preprintnumbers,superscriptaddress,floatfix]{revtex4-2}

\usepackage{epsfig}
\usepackage{amsmath}
\usepackage{amssymb}
\usepackage{latexsym}
\usepackage{xspace}
\usepackage[colorlinks=true,linktocpage=true,linkcolor=blue,citecolor=blue,allcolors=blue]{hyperref}
\usepackage{url}
\usepackage[utf8]{inputenc}
\usepackage{indentfirst}
\usepackage{enumerate}
\usepackage{color}
\usepackage{relsize}
\usepackage{graphicx}
\usepackage{dcolumn}
\usepackage{bm}
\usepackage[caption=false,position=top]{subfig}
\usepackage[english]{babel}
\usepackage{url}
\usepackage[colorlinks=true,linktocpage=true,linkcolor=blue,citecolor=blue,allcolors=blue]{hyperref}

\newcommand{\be}{\begin{equation}}
\newcommand{\ee}{\end{equation}}

\begin{document}

\title{Thermal Axion Production at Low Temperatures: \\ A Smooth Treatment of the QCD Phase Transition}

\author{Francesco D'Eramo}
 \email{francesco.deramo@pd.infn.it}
\author{Fazlollah Hajkarim}
 \email{fazlollah.hajkarim@pd.infn.it}
 \author{Seokhoon Yun}
 \email{seokhoon.yun@pd.infn.it}

\affiliation{Dipartimento di Fisica e Astronomia, Universit\`a degli Studi di Padova, Via Marzolo 8, 35131 Padova, Italy}
\affiliation{INFN, Sezione di Padova, Via Marzolo 8, 35131 Padova, Italy}

\date{August 9, 2021}

\begin{abstract}

We study thermal axion production around the confinement scale. At higher temperatures, we extend current calculations to account for the masses of heavy quarks, whereas we quantify production via hadron scattering at lower temperatures. Matching our results between the two opposite regimes provides us with a continuous axion production rate across the QCD phase transition. We employ such a rate to quantify the axion contribution to the effective number of neutrino species.

\end{abstract}

\maketitle


\noindent {\bf Introduction.} The early Universe served as the stage for numerous fascinating phenomena, and they are all hidden behind the curtain of the last scattering surface. A few of them could leave cosmological imprints that we search for in our detectors today. Dark radiation is a notable example: relativistic relics produced at early times, whether they achieve thermalization with the primordial bath or not, at some point start to free-stream unhindered until the present time. If they are still relativistic at recombination, they contribute to the radiation energy density and alter the cosmic microwave background (CMB) anisotropy spectrum at small angular scales. Historically, this effect is quantified in terms of an additional number of effective neutrino species $\Delta N_{\rm eff}$. The Planck Collaboration~\cite{Aghanim:2018eyx} places the constraint $N_{\rm eff} = 2.99 \pm 0.17$~\footnote{The number of effective neutrino species is the sum of the standard model contribution and the new physics one, $N_{\rm eff} = N^{\rm SM}_{\rm eff} + \Delta N_{\rm eff}$, with the former $N^{\rm SM}_{\rm eff} \simeq 3.0440$ due to non instantaneous neutrino decoupling~\cite{Mangano:2001iu,Bennett:2019ewm,Akita:2020szl,Bennett:2020zkv}.}. 

Future CMB-S4 surveys constitute a timely and powerful probe for light and elusive new physics beyond the standard model (SM)~\cite{Brust:2013xpv,Baumann:2016wac}. The projected reach, $\Delta N_{\rm eff}^{{\rm CMB-S4}}(1 \sigma) \simeq 0.02 - 0.03$~\cite{CMB-S4:2016ple,Abazajian:2019eic}, will allow us to detect relics that decoupled above the electroweak phase transition. For some candidates, CMB will provide information that is complementary to other search strategies, while for others it could be the only diagnostic. 

The QCD axion~\cite{Wilczek:1977pj,Weinberg:1977ma} is a strongly motivated candidate for physics beyond the SM. It solves the strong {\it CP} problem via the Peccei-Quinn (PQ) mechanism~\cite{Peccei:1977np,Peccei:1977hh}, it is a viable dark matter candidate~\cite{Preskill:1982cy,Abbott:1982af,Dine:1982ah} and responsible for intriguing new early Universe dynamics~\cite{Marsh:2015xka}, and it is the target of a multitude of experiments~\cite{Graham:2015ouw,Irastorza:2018dyq,Sikivie:2020zpn}. The production of axion dark radiation in the early Universe and its cosmological imprint on the CMB is no exception~\footnote{This cosmic axion background could be detected even with terrestrial experiments, see the recent Ref.~\cite{Dror:2021nyr}.}, and the astonishing future discovery reach makes rigorous theoretical predictions a top priority.

Early Universe axion dynamics is challenging on several levels. Quarks and gluons are the relevant strongly interacting degrees of freedom at high energy, and the axion has an anomalous coupling with gluons plus model-dependent interactions with quarks~\cite{Kim:2008hd,DiLuzio:2020wdo}. This picture breaks down below the proton mass where strong interactions become nonperturbative and quarks are confined within hadrons. Quantifying the production rate across the QCD phase transition (QCDPT) in the strongly coupled regime is far from being straightforward, and axion production was analyzed either exclusively above the confinement scale~\cite{Turner:1986tb,Masso:2002np,Graf:2010tv,Salvio:2013iaa,Ferreira:2018vjj,Arias-Aragon:2020qtn,Arias-Aragon:2020shv} or below~\cite{Berezhiani:1992rk,Chang:1993gm,Hannestad:2005df,DEramo:2014urw,Kawasaki:2015ofa,Ferreira:2020bpb,DiLuzio:2021vjd,Carenza:2021ebx}.
 
In this Letter, we provide for the first time a smooth and continuous result for the axion production rate across the QCDPT. We extend previous rate calculations above and below the confinement scale in the regimes where we have perturbative control, and we interpolate our results in the intermediate region. As a result, we find a continuous rate for every value of the bath temperature $T$ across the QCDPT. As we will see explicitly, such an interpolation is needed to exploit results from CMB-S4 experiments. We focus on the axion coupling to gluons 
\be
\mathcal{L}_{aG} = \frac{\alpha_s}{8 \pi} \frac{a}{f_a} G^A_{\mu\nu} \widetilde{G}^{A \mu\nu} \ .
\label{eq:LagAxion}
\ee
We denote the axion field and its decay constant by $a$ and $f_a$, respectively, and the strong fine structure constant in terms of the QCD coupling $g_s$ by $\alpha_s = g_s^2 / (4 \pi)$. The gluon field strength $G^A_{\mu\nu}$ and its dual $\widetilde{G}^{A \mu\nu} \equiv \epsilon^{\mu\nu\rho\sigma} G^A_{\rho\sigma} / 2$ have both $SU(3)_c$ adjoint color indices $A=1,2,\dots,8$. The operator in Eq.~\eqref{eq:LagAxion} must be present in any PQ theory to solve the strong {\it CP} problem, and in this respect, this is a model-independent axion interaction. Besides axion self-interactions that do not impact our study, it is the only relevant axion interaction above the QCDPT within the Kim-Shifman-Vainshtein-Zakharov (KSVZ)~\cite{Kim:1979if,Shifman:1979if} framework where it is the radiative remnant at low energy of a heavy, colored, and PQ charged fermion.

The zero temperature axion mass results in~\cite{Bardeen:1978nq,GrillidiCortona:2015jxo,Gorghetto:2018ocs} 
\be
m_a \simeq \frac{\sqrt{m_u m_d}}{m_u + m_d} \frac{m_\pi f_\pi}{f_a} \ ,
\label{eq:mavsfa}
\ee
where $m_{u,d}$ are the light quark masses, $m_\pi$ the pion mass and $f_\pi$ the pion decay constant~\footnote{We adopt the conventions of Ref.~\cite{GrillidiCortona:2015jxo} with numerical value of the pion decay constant $f_\pi = 92(1) \, {\rm MeV}$.}. In the most conservative scenario with only the interaction in Eq.~\eqref{eq:LagAxion} switched on, the best bound on $f_a$ (and, consequently, on the axion mass $m_a$) comes from stellar cooling arguments~\cite{Fischer:2016cyd,Chang:2018rso,Carenza:2019pxu}. However, this bound relies strongly upon numerical simulations of supernova explosions that in turn demand several assumptions, and it is not a statistically rigorous bound. Future CMB-S4 surveys have the potential of putting  the axion mass bound in this region on solid ground, and this requires a continuous rate across the QCDPT. 

First, we consider thermal gluon scattering above the QCDPT and we regularize IR divergences due to the long-range gluon mediated interactions. We extend current treatments that are valid only above the electroweak phase transition. Below the confinement scale, we account for axion production via pion scatterings mediated by low energy interactions generated from Eq.~\eqref{eq:LagAxion}. The QCDPT at vanishing baryon chemical potential is a smooth crossover where thermodynamic variables are continuous~\cite{Aoki:2006we,HotQCD:2014kol}. Thus we join and interpolate the axion production rate across this threshold, and we provide a continuous result that is valid at all temperatures~\footnote{Ref.~\cite{Venumadhav:2015pla} applied the same method to compute the sterile neutrino production rate across the QCDPT.}. We conclude with the evaluation of the axion dark radiation amount quantified by $\Delta N_{\rm eff}$. Technical details about our analysis as well as additional calculations within UV complete axion models can be found in Ref.~\cite{DEramo:2021lgb}. 

\vspace{0.1cm}
\noindent {\bf Rate above QCDPT.} Before strong interactions confine, quark and gluon scatterings produce axions. Famously, the expansion parameter in thermal field theory is $g_s$ [rather than $\alpha_s / (4\pi)$] as a consequence of collinear enhancements~\cite{Braaten:1991dd}. As we approach the QCDPT, the strong coupling constant $g_s$ grows and we may need to resum processes involving many particles. Luckily, such an enhancement is absent for the specific operator in Eq.~\eqref{eq:LagAxion} and we can restrict ourselves to binary collisions~\footnote{This was first pointed out by Reference.~\cite{Salvio:2013iaa}.}.

The relevant processes producing axions in this regime are gluon scatterings ($g+g \rightarrow g+a$), quark and antiquark annihilations ($q+\bar{q} \rightarrow g+a$), and Primakoff-like scatterings ($q/\bar{q}+g \rightarrow q/\bar{q}+a$). Crucially, the exchange of massless gluons lead to IR divergences that need to be taken care of. The early analysis in Ref.~\cite{Masso:2002np} regularized such an unpleasant IR behavior with an explicit gluon Debye mass. Later on, Ref.~\cite{Graf:2010tv} employed the more suitable thermal field theory formalism, and analyzed the problem in the so-called hard thermal loop (HTL) region ($g_s \ll 1$). Reference~\cite{Salvio:2013iaa} extended this treatment beyond the HTL region but only at high temperatures, $T \gtrsim 10^4 \, {\rm GeV}$. We need to extend this analysis to lower temperatures.

The imaginary part of the axion self-energy $\Pi_a$ controls the production rate via the relation~\cite{Weldon:1990iw,Gale:1990pn}
\be
\gamma_a \equiv \frac{d N_a}{dV dt} = - 2\int \frac{d^3 p_a}{2 E_a \left(2\pi\right)^3} f_{\rm BE}(E_a)\,{\rm Im} \,\Pi_a 
\label{eq:RateAboveQCDPT}
\ee
with $f_{\rm BE}(E_a)$ the Bose-Einstein distribution evaluated at $E_a = \left|\vec{p}_a\right|$. All we need is the axion two-point function sourced by the cubic vertex in Eq.~\eqref{eq:LagAxion}. The only diagram contributing to the rate is the one loop axion two-point function with virtual gluons exchanged, and with the tree-level gluon propagator replaced with the resummed thermal one. Reference~\cite{Salvio:2013iaa} dubbed this the ``decay'' diagram because it describes thermal gluon decays that become available thanks to finite temperature corrections. We work at the leading order in the strong coupling constant $g_s$ since we consider binary collisions. 

Reference~\cite{Rychkov:2007uq} provides general expressions for the resummed gluon propagator at one-loop, and we account for both virtual gluons and quarks. If one wants to determine the production rate just above the QCDPT, the decoupling of heavy quarks at their mass thresholds must be done properly. For this reason, and unlike previous treatments in the literature, we keep the quark masses finite in our analytical expressions. With the resummed propagator in hand, we express the axion self-energy $\Pi_a$ appearing in the rate in Eq.~\eqref{eq:RateAboveQCDPT} in terms of the longitudinal and transverse gluon spectral densities~\cite{Bellac:2011kqa,Laine:2016hma}. The final expression involves several numerical integrations, and we adopt the strategy of Ref.~\cite{Rychkov:2007uq} that divides the spectral densities into pole and continuum parts to deal with numerical difficulties. We employ the `\texttt{RunDec}'~\cite{Chetyrkin:2000yt,Herren:2017osy} code to account for the running of the $\alpha_s$ up to four loops. 

\begin{figure}
\centering
\includegraphics[width=0.45\textwidth]{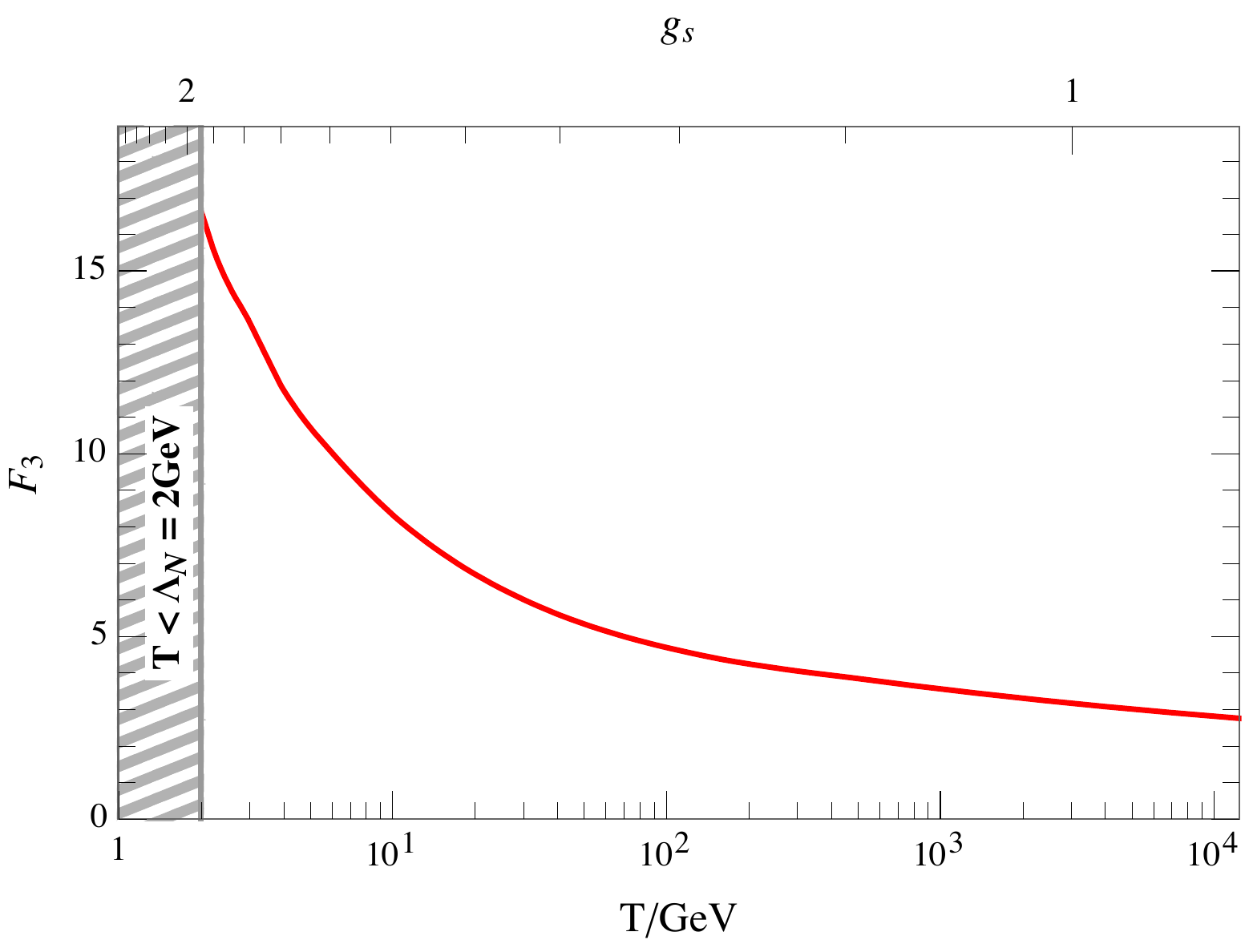}
\caption{Function $F_3(T)$, which controls the rate via Eq.~\eqref{eq:F3function}, as a function of the temperature toward the QCDPT.}
\label{fig:F3}
\end{figure}

We follow Ref.~\cite{Salvio:2013iaa} and parametrize the rate 
\be
\gamma_{gg} \equiv \frac{d N_a}{dV dt} = \frac{2 \zeta (3) d_g}{\pi^3}\left(\frac{\alpha_s }{8\pi f_a}\right)^2 F_3\left(T\right) \, T^6  
\label{eq:F3function}
\ee
with the Riemann zeta function $\zeta(3) \simeq 1.2$ and $d_g = 8$ the dimension of the adjoint representation of the $SU(3)$ color gauge group. The temperature dependence via the control function $F_3(T)$ is shown in Fig.~\ref{fig:F3}. 

\vspace{0.1cm}
\noindent {\bf Rate below QCDPT.} At temperatures below the confinement scale, axion production is controlled by hadron scatterings. The rate computation must resort to nonperturbative techniques, and we employ the ones of chiral perturbation theory (ChPT) in this regime. Large instanton effects make the operator in Eq.~\eqref{eq:LagAxion} rather inconvenient to identify axion couplings. We rotate it away through an axion-dependent anomalous field redefinition of the light quarks, $q \rightarrow \exp[ -i c_q (a / f_a) \gamma^5] q$, with $c_q = M_q^{-1}/2 \, {\rm Tr} [M_q^{-1}]$ and $M_q = {\rm diag} (m_u,m_d,m_s)$. While we remove the coupling to gluons, axion derivative interactions to quark currents appear in the Lagrangian
\be
\mathcal{L}_{aq} = \frac{\partial_\mu a}{f_a} \sum_{q=u,d,s} c_q \, \bar{q}\gamma^\mu\gamma^5 q \,.
\label{eq:LagAxion2}
\ee
We match this operator onto a low-energy effective theory with only hadrons in the spectrum. As we discuss shortly, we apply this theory only up to energies $\Lambda_{\rm ChPT} \sim \mathcal{O}(100 \, {\rm MeV})$. Thus the leading contribution to the production rate comes from pion scatterings, and the ones from processes involving baryons (e.g., nucleons) and heavy mesons (e.g., $K$ and $\eta$) are highly suppressed. 

Axion-pion interactions at low energy are an application of the general formalism provided by Refs.~\cite{Gasser:1983yg,Gasser:1984gg}. We match spin-one standard model currents with the same symmetry properties between the UV and IR, with the axion acting as a spectator in this operation. We find the low-energy Lagrangian~\cite{Srednicki:1985xd,Georgi:1986df,GrillidiCortona:2015jxo}
\be
 \mathcal{L}_{a\pi} = \frac{\partial_\mu a}{f_a} \frac{c_{a\pi\pi\pi}}{f_\pi} \, J^\mu_\pi  \ ,
\ee 
where we introduce the spin-one current made of pions $J^\mu_\pi =  \left[ \pi^0\pi^+\partial^\mu \pi^- + \pi^0\pi^-\partial^\mu \pi^+  - 2 \pi^+\pi^-\partial^\mu \pi^0 \right]$, and the dimensionless coefficient $c_{a\pi\pi\pi}$ explicitly reads
\be
c_{a\pi\pi\pi} = \frac{2}{3}{\rm Tr}\left[\lambda^3 c_q\right] = \frac{1}{3}\frac{m_u^{-1} - m_d^{-1}}{m_u^{-1}+m_d^{-1}+m_s^{-1}} 
\ee 
with $\lambda^3$ a Gell-Mann matrix. We use the quark mass ratios values $m_u/m_d = 0.48(3)$ and $m_u/m_s = 0.024(1)$~\footnote{We follow Ref.~\cite{GrillidiCortona:2015jxo} and average over the values given in Refs.~\cite{deDivitiis:2013xla,Horsley:2015eaa,MILC:2015ypt}. We use the central values of $c_{a\pi\pi}$ and $f_\pi$ in our analysis, and the experimental uncertainties on their values do not affect our predictions for $\Delta N_{\rm eff}$.}, and they give the coupling $c_{a\pi\pi\pi} \simeq 0.12(1)$. 

These operators mediate axion production via the pion binary collisions $\pi^+ + \pi^-\rightarrow \pi^0 + a$, $\pi^+ + \pi^0 \rightarrow \pi^+ + a$, and $\pi^- + \pi^0 \rightarrow \pi^- + a$. We compute the scattering cross section with the aid of ``{\tt FeynCalc}"~\cite{Shtabovenko:2016sxi,Shtabovenko:2020gxv}, and we derive the axion production rate in this regime
\be
\gamma_{\pi \pi} = \sum_{\pi \, {\rm scatterings}} n_{\pi_i}^{\rm eq} n_{\pi_j}^{\rm eq} \langle \sigma_{\pi_i \pi_j \rightarrow \pi_k a} v_{\rm rel} \rangle \ .
\label{eq:gammapipi}
\ee
The sum runs over the production processes, and $n_{\pi_i}^{\rm eq}$ denotes the pion equilibrium number density. Each cross section times the M{\o}ller velocity $v_{\rm rel}$ is thermally averaged over all the possible initial state configurations. 

\vspace{0.1cm}
\noindent {\bf Matching across the QCDPT.} The axion production rate above the confinement scale is quantified by the function $\gamma_{gg}$ in Eq.~\eqref{eq:F3function}, with the function $F_3(T)$ in Fig.~\ref{fig:F3}. The validity of this result extends down to a temperature $T \simeq \Lambda_N \simeq 2 \, {\rm GeV}$ where the strong coupling constant is $\alpha_s(\Lambda_N) \simeq 0.3$. In the opposite regime, below the confinement scale, the rate is driven by pion scattering through the function $\gamma_{\pi \pi}$ defined in Eq.~\eqref{eq:gammapipi}. Its evaluation relies upon the ChPT framework which is valid up to a UV cutoff $\Lambda_{\rm ChPT}$ where such a formalism breaks down. 

Up to what temperatures are we allowed to push the validity of $\gamma_{\pi \pi}$? We treat the primordial plasma within the hadron resonance gas (HRG) formalism~\cite{Hagedorn:1984hz,Huovinen:2009yb,Megias:2012hk}, and lattice QCD results show how this is inconsistent above $T>150\,{\rm MeV}$~\cite{Venumadhav:2015pla}. Besides our treatment of the thermal bath, the analysis in Ref.~\cite{DiLuzio:2021vjd} pointed out how one loses perturbativity control above $T>62\,{\rm MeV}$ if the leading order ChPT is employed to compute axion production.

\begin{figure}
\centering
\includegraphics[width=0.45\textwidth]{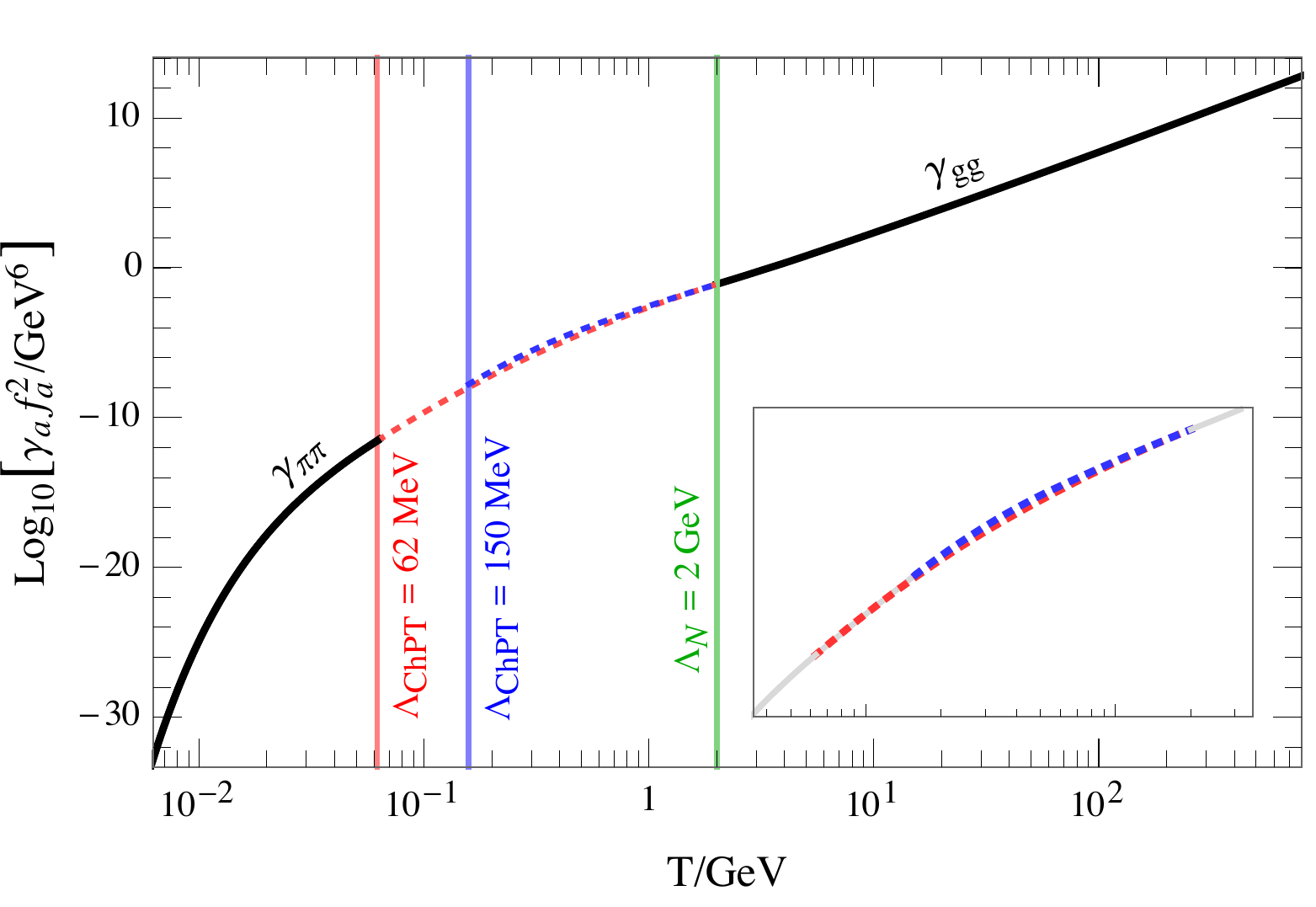}
\caption{Axion production rate across the QCDPT. At high temperatures ($T > \Lambda_N$), the production is driven by thermal gluon scatterings whereas pion binary collisions are the main source of axions at low temperatures ($T < \Lambda_{\rm ChPT}$). We interpolate between the two regimes (see text for details).}
\label{fig:rate}
\end{figure}

Both values of this UV cutoff $\Lambda_{\rm ChPT}$ for the function $\gamma_{\pi\pi}$ are smaller than $\Lambda_N$. We interpolate the axion production rate between $\Lambda_{\rm ChPT}$ and $\Lambda_N$ with the ``spline'' (cubic) method. We account for both options for $\Lambda_{\rm ChPT}$, with results shown in Fig.~\ref{fig:rate}. Here, the red-dashed and blue-dashed lines correspond to the interpolation with the choices $\Lambda_{\rm ChPT} =  62$ and $150\,{\rm MeV}$, respectively. As it is manifest from the figure, the two lines coincide with each other. The insensitiveness of the matching result to the detailed choice for $\Lambda_{\rm ChPT}$ is reasonable since the QCDPT is a smooth crossover without any discontinuity in the thermodynamic variables~\footnote{We discuss in Ref.~\cite{DEramo:2021lgb} theoretical uncertainties on $\Delta N_{\rm eff}$ due to the interpolation procedure and show how our predictions are robust.}.

\vspace{0.1cm}
\noindent {\bf Axion dark radiation.} The axion number density $n_a$ evolves according to the Boltzmann equation
\be
\frac{d n_a}{d t} + 3 H n_a = \gamma_a \left( 1 - \frac{n_a}{n_a^{\rm eq}} \right) \ .
\label{eq:BE1}
\ee 
The term proportional to the Hubble rate on the left-hand side accounts for the dilution due to the expansion. Collisions are responsible for the right-hand side. The term with the equilibrium axion number density $n_a^{\rm eq}$ describes the inverse (axion destruction) process, and the production rate $\gamma_a$ is the quantity shown in Fig.~\ref{fig:rate}.

Dimensionless variables are convenient for numerical solutions so we employ the axion comoving number density $Y_a = n_a / s$. The entropy density is a function of the bath temperature $T$, $s = 2 \pi^2 g_{*s}(T)T^3/45$, with $g_{*s}(T)$ the effective entropic degrees of freedom. Concerning the evolution variable, we choose the combination $x = M / T$, with the choice of $M$ purely conventional; throughout this analysis we set $M = 1 \, {\rm GeV}$ since we are investigating axion production around the QCDPT. The Boltzmann equation in terms of these new variables reads
\be
\frac{d Y_a}{d \log x} = \left( 1 - \frac{1}{3} \frac{d \log g_{*{s}}}{d \log x} \right) \frac{\gamma_a(x)}{H(x) s(x)} \left( 1 - \frac{Y_a}{Y_a^{\rm eq}} \right) \ . 
\label{eq:BE2}
\ee
Solving this equation requires knowing the composition of the primordial bath. Besides the already discussed entropic degrees of freedom $g_{*s}(T)$, we also need the ones [$g_{*}(T)$] contributing to the energy density since they affect the expansion rate~\footnote{The radiation bath energy density scales with the temperature as $\rho = (\pi^2/30) g_*(T) T^4$. The Hubble rate follows from the Friedmann equation, $H(T) = \sqrt{\rho} / (\sqrt{3} M_{\rm Pl})$, where $M_{\rm Pl}$ is the reduced Planck mass.}. 

We solve this differential equation numerically starting from an initial temperature $T_i$. We consider two opposite initial conditions: vanishing initial density and a thermal axion population in equilibrium with the plasma at $T_i$. These two extremes cover the broad spectrum of possibilities such as axion early thermalization from inflaton decays. Nevertheless, the prediction for $\Delta N_{\rm eff}$ in the region where the signal is detectable is not sensitive to the initial conditions unless we consider large values of $f_a$.

As the Universe expands and cools down, and regardless of the details of axion production, there is a temperature below which the axion comoving density freezes to a constant value $Y_a^\infty = {\rm const}$. This can happen either because the bath particles participating in the production processes become nonrelativistic and their number density gets exponentially suppressed, or because the Universe gets too cold and diluted to have significant collisions within a Hubble time. Such an asymptotic value corresponds to $\Delta N_{\rm eff} \simeq 75.6 \, (Y_a^\infty)^{4/3}$.

\vspace{0.1cm}
\noindent {\bf Outlook.} The QCD axion is one of the most motivated hypothetical particles for physics beyond the SM: it solves the strong {\it CP} problem via the PQ mechanism, and it is a viable dark matter candidate. Extraordinary efforts from experiments, which have been literally blossoming in the last decade with novel ideas, make the present time remarkably exciting for axion physics. 

Hot axions produced in the early Universe leave a cosmological imprint through $\Delta N_{\rm eff}$. Intriguingly, such a population of relativistic axions could lie at the origin of the discrepancy between high- and low-redshift measurements of the Hubble expansion rate~\cite{DEramo:2018vss}. The impressive projections by future CMB-S4 surveys make this a central signature of PQ theories that is complementary to other search strategies. Reliable theoretical predictions quantifying this effect are of paramount importance.

In this work, we filled an important gap: the evaluation of the axion production rate across the QCDPT. Focusing on the model-independent axion interaction in Eq.~\eqref{eq:LagAxion}, we computed the production rate above and below the confinement scale, and we provided a smooth interpolation between these two regimes as shown in Fig.~\ref{fig:rate}. 

\begin{figure}
\centering
\includegraphics[width=0.45\textwidth]{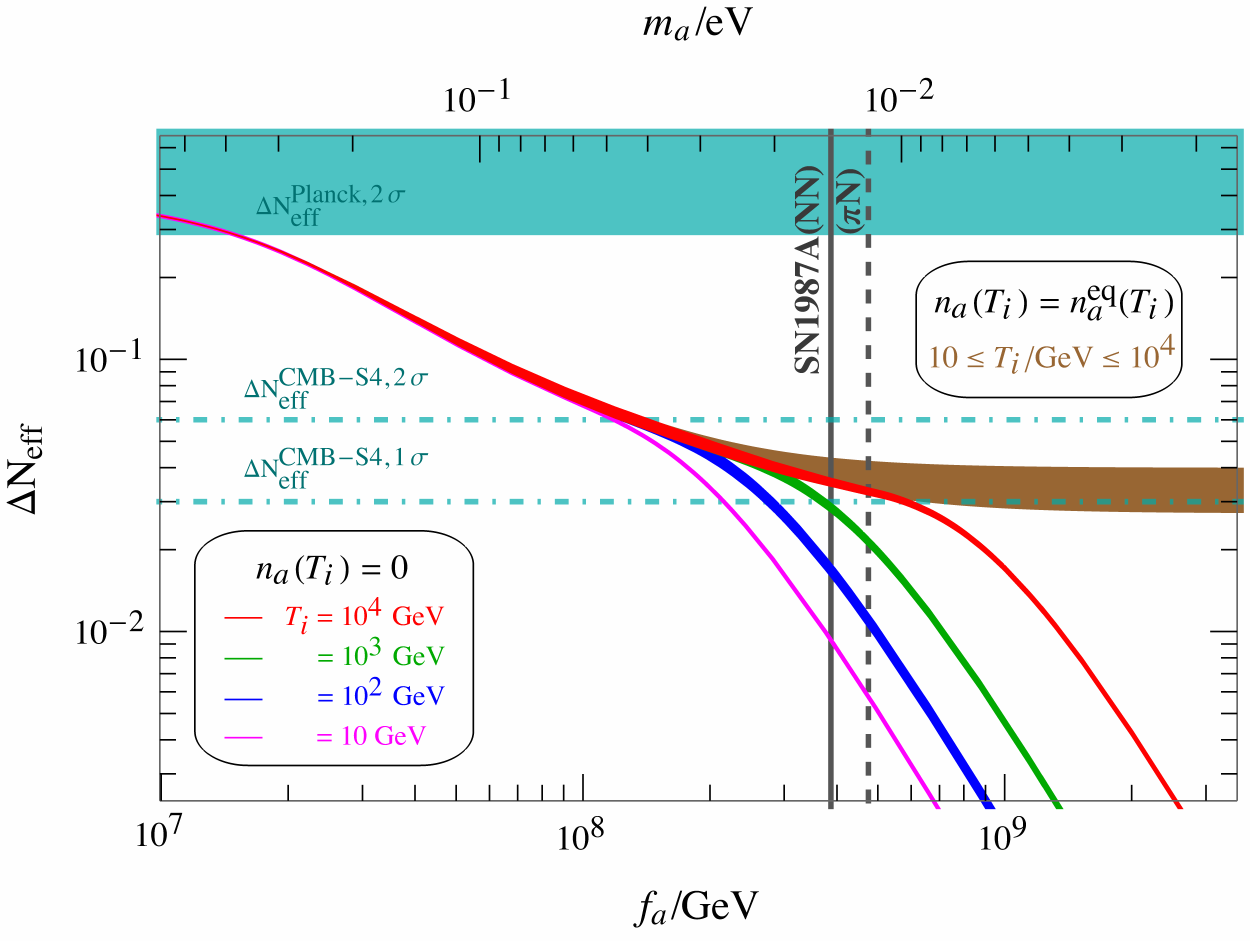}
\caption{$\Delta N_{\rm eff}$ from axion dark radiation as a function of the axion decay constant $f_a$. We consider different values for the initial temperature $T_i$, and we account for both a vanishing initial density as well as an initial thermal axion population.}
\label{fig:results}
\end{figure}

The central result of our analysis is in Fig.~\ref{fig:results} where we show the amount of axion dark radiation as a function of the axion decay constant $f_a$. Vertical lines identify the SN1987A bound on $f_a$. Equivalently, this is a bound on the axion mass $m_a$ as can be understood with the help of Eq.~\eqref{eq:mavsfa} and as illustrated explicitly in the upper vertical axis. The solid vertical line indicates the bound $f_a \gtrsim 3.87 \times 10^8 \, {\rm GeV}$ that was obtained by Ref.~\cite{Carenza:2019pxu} accounting for nucleon bremsstrahlung. Recently, it has been argued that (negative-charged) pions might be abundant during supernovae explosions~\cite{Fore:2019wib}, and, consequently, the $\pi^- + p \rightarrow n + a$ process could contribute to the axion luminosity~\cite{Carenza:2020cis,Fischer:2021jfm,Choi:2021ign} significantly. This implies the bound $f_a \gtrsim 4.75 \times 10^8 \, {\rm GeV}$~\cite{Carenza:2020cis} that we show as a dashed vertical line. Furthermore, we shade away the region excluded by the Planck data~\cite{Aghanim:2018eyx} (green). Solid continuous lines correspond to the predicted $\Delta N_{\rm eff}$ when we solve the Boltzmann equation with vanishing initial axion density starting from the initial temperature $T_i$. We employ for the Boltzmann evolution the effective relativistic degrees of freedom provided in Refs.~\cite{Drees:2015exa,Saikawa:2018rcs}, and we shade the region in between the two solutions. Our predictions do not depend on this choice. The brown band  is the predicted $\Delta N_{\rm eff}$ when we begin the evolution with a full thermal equilibrium population. As expected, the difference between these two initial conditions arises at large $f_a$ when it is harder to thermalize if we begin with vanishing initial density.  Current CMB experiments, such as the Planck satellite, are testing a region in conflict with stellar bounds for the hadronic axion interaction in Eq.~\eqref{eq:LagAxion}. Future CMB-S4 surveys will probe values of $f_a$ as large as $10^9 \, {\rm GeV}$ even if the primordial bath begins its existence after inflation with no axions. 

Our analysis paves the way for several future directions, and we conclude by mentioning two possibilities. On one hand, the interaction in Eq.~\eqref{eq:LagAxion}, although model independent, is far from being the only axion coupling within concrete models~\cite{DiLuzio:2020wdo}. One can predict the amount of axion dark radiation for all models available in the literature building upon the work that we have presented here, and potentially use a future detection of $\Delta N_{\rm eff}$ as a discriminant. The recent work in Ref.~\cite{DEramo:2021lgb} provided a prediction for $\Delta N_{\rm eff}$ for the DFSZ axion~\cite{Zhitnitsky:1980tq,Dine:1981rt}, other motivated frameworks are flavor-violating axion models~\footnote{Flavor-violating couplings can arise from radiative corrections~\cite{Choi:2017gpf,Chala:2020wvs,Bauer:2020jbp,Choi:2021kuy,Bonilla:2021ufe} and/or PQ charge assignments~\cite{Ema:2016ops,Calibbi:2016hwq}.} where the production is controlled by bath particle decays. On the other hand, one can investigate axion production mediated by the operator in Eq.~\eqref{eq:LagAxion}, with rate shown in Fig.~\ref{fig:rate}, for modified cosmological histories. 

\vspace{0.3cm}
\noindent {\it Acknowledgements.} Authors acknowledge L. Di Luzio, W. Fischler, S. H. Lim, S. D. McDermott, A. Mirizzi, J. Schaffner-Bielich, C. S. Shin, and L. Tolos for useful discussions. This work is supported by the research grants: ``The Dark Universe: A Synergic Multi-messenger Approach'' No. 2017X7X85K under the program PRIN 2017 funded by the Ministero dell'Istruzione, Universit\`a e della Ricerca (MIUR); ``New Theoretical Tools for Axion Cosmology'' under the Supporting TAlent in ReSearch@University of Padova (STARS@UNIPD). The authors also supported by Istituto Nazionale di Fisica Nucleare (INFN) through the Theoretical Astroparticle Physics (TAsP) project. F.D. acknowledges support from the European Union's Horizon 2020 research and innovation programme under the Marie Sk\l odowska-Curie Grant Agreement No. 860881-HIDDeN. 

\bibliographystyle{apsrev4-2}
\bibliography{HotAxionsQCD}

\end{document}